\documentstyle[preprint,aps,eqsecnum,epsfig]{revtex}
\tighten
\begin{document}
\draft
\title{\bf SELF-CONSISTENT DYNAMICS OF INFLATIONARY PHASE TRANSITIONS \footnote{To appear in the Proceedings of SEWM'97}}
\author{{\bf D. Boyanovsky$^{(a)}$, D. Cormier$^{(b)}$,
H.J. de Vega$^{(c)}$, R. Holman$^{(b)}$, S. P. Kumar$^{(b)}$}}
\address
{ (a)  Department of Physics and Astronomy, University of
Pittsburgh, Pittsburgh, PA. 15260, U.S.A. \\
 (b) Department of Physics, Carnegie Mellon University, Pittsburgh,
PA. 15213, U. S. A. \\
 (c)  LPTHE, \footnote{Laboratoire Associ\'{e} au CNRS UA280.}
Universit\'e Pierre et Marie Curie (Paris VI) 
et Denis Diderot  (Paris VII), Tour 16, 1er. \'etage, 4, Place Jussieu
75252 Paris, Cedex 05, France }
\date{July 1997}
\maketitle
\begin{abstract}
The physics of the inflationary universe requires  the study of the out of
equilibrium evolution of quantum fields in curved spacetime. 
We present the evolution for both the geometry and the matter
(described by the quantum inflaton field) by means of the  
non-perturbative large $ N $ limit combined with 
semi-classical gravitational dynamics including the back-reaction of
quantum fluctuations self-consistently for a new inflation scenario.
 We provide a criterion for the
validity of the classical approximation and a full analysis of the
case in which spinodal quantum fluctuations drive the evolution of the
scale factor. Under carefully determined conditions, we show that the full 
field equations may be well approximated by those of a single composite 
field which obeys the classical equation of motion in all cases.
  The de Sitter stage is found to be followed by a matter dominated phase.
We compute the spectrum of scalar density perturbations and argue that
the spinodal instabilities are responsible for a `red' spectrum with 
more power at longer wavelengths. A criterion for the validity of these
models is provided and contact with the reconstruction program is established.

\end{abstract}

\section{Introduction and Motivation}
\subsection{Preliminaries}
During the last decade Particle Physics Cosmology matured from the
realm of speculation to that of testable predictability. The next
generation of satellite and balloon borne experiments will provide an
unprecedented test of the fundamental ideas of Early Universe 
Cosmology and will validate or rule out current theories of structure
formation and other aspects of the cosmology of the early universe\cite{kolb,turner}. 
Standard Big Bang cosmology is based on a homogeneous and isotropic
expanding universe and is confirmed by observations: i) redshifts of
objects far away have been measured and confirm Hubble's law of expansion,
ii) the cosmic microwave background radiation (CMBR) has been measured
to be an almost perfect blackbody with temperature $T= 2.728 \pm 0.002K$ 
and is a remnant of the early hot and dense stages of evolution 
after the Big Bang, iii) light element abundance of $D, ^3He, ^4He, ^7Li$
which is consistent with what is predicted by Nucleosynthesis. 

 The CMBR gives
evidence of homogeneity and isotropy on scales larger than about 100 Mpc,
with small temperature fluctuations $\Delta T/T < 10^{-5}$ on angular
scales that range from $1^o$ to about $90^o$\cite{smoot,white}. This high
degree of homogeneity and isotropy on large scales presents one of the
important puzzles of standard big bang cosmology, i.e. the horizon
problem. 

Distances in a spatially flat, homogeneous and isotropic cosmology 
are measured with the FRW metric
\begin{equation}
ds^2 = dt^2-a^2(t)d{\vec x}^2, \label{FRW}
\end{equation}
with $t$ the comoving or cosmic time and $a(t)$ the scale factor. Physical
distances along null geodesics determine the limit of causal correlations.
These are given by
\begin{equation}
d_h(t)= a(t) \int^t \frac{dt'}{a(t')}. \label{horizon}
\end{equation}
this distance determines the causal horizon, events that are separated
by distances larger than $d_h(t)$ cannot be causally correlated  
by microphysical processes. Another important physical scale is the
Hubble radius $d_H(t) = H^{-1}(t)= (\dot{a}(t)/a(t))^{-1}$. In the cases
of interest, $d_h(t)$ and $d_H(t)$ are proportional and we will call 
$d_H(t)$ the horizon indistinguishably.  Let us consider a physical
distance today $\lambda(t_0)$ over which the CMBR determines
homogeneity and isotropy to one part in $10^5$. At a time $t$ earlier,
the size of 
this patch is given by $\lambda(t) = \lambda(t_0)a(t)$ (where we have
chosen $a(t_0)=1$). For a matter or radiation dominated universe (most
of the life of the Universe), $a(t) \propto t^n$ ($n=1/2$ for radiation,
$n=2/3$ for matter domination). The horizon size (or Hubble radius) is
given by $d_H(t) \propto t$ and the ratio $\lambda(t)/d_H(t)$ increases
as we evolve back to an earlier epoch. Thus at some time in the far 
past, the scale
$\lambda(t_0)$ will become larger than the horizon. In particular, if
$\lambda(t_0) \approx 100 \mbox{ Mpc }$ this scale will be larger than the
horizon near the time of decoupling of matter and radiation,
 at red-shifts $z \approx 1100$, and therefore it is extremely difficult
to explain how such a region which was causally disconnected at
decoupling can be so homogeneous and isotropic. This is the essence of
the horizon problem\cite{kolb,turner}.
 
Inflation was proposed over
a decade ago to solve long standing problems of Standard Big Bang
cosmology such as the homogeneity and horizon
problems\cite{guth,linde,andy}. Inflation corresponds to a stage of
exponential expansion of the Universe during which 
the Hubble radius (horizon) remains constant and physical scales grow
exponentially. Thus, in this scenario, microphysical scales  smaller
than the horizon  over which
causal processes establish correlations cross the horizon during
the inflationary stage. After the inflationary stage and through 
particle physics processes (reheating), the Universe becomes radiation
and matter dominated and the scales that had crossed outside the
Hubble radius  
during the inflationary stage, re-enter the horizon. Perturbations
on these scales then grow under gravitational (Jeans) instability to
form the large scale structures that we see today. Thus the inflationary
proposal solves the homogeneity and horizon problems by allowing
scales to cross the horizon {\em twice}, from being subhorizon
initially  becoming superhorizon during inflation, and re-entering
the horizon during the period of radiation or matter domination. 

Quantum field theory combined with particle physics models provides the
framework to implement the inflationary idea. For an inflationary
stage the equation of state has to be dominated by a term similar to a
cosmological 
constant leading to a negative pressure. Through the
Einstein-Friedmann equations, this yields 
to an exponential expansion of the scale
factor. There are many different inflationary scenarios: `old', `new',
`natural', `hybrid' etc.\cite{kolb,turner}, 
but despite the plethora of different implementations of the original
idea, there are some robust features of inflation common to practically
all models:
i) Inflation predicts a {\em flat} universe, i.e. the total energy
density at all times is the critical energy density, ii) an almost
scale invariant spectrum of scalar density perturbations, with amplitudes
that are bound by the CMBR inhomogeneities $\Delta T/T$, 
iii) approximately scale invariant spectrum of gravitational waves
(tensor perturbations). The next generation of satellites (MAP/PLANCK)
will provide more stringent bounds on temperature inhomogeneities and
a firmer determination of the spectrum of scalar and tensor perturbations,
and the next generation of gravitational wave detectors (LIGO, VIRGO,
LISA) may detect the gravitational waves and their
spectrum. Determinations of the spectrum of scalar and tensor
perturbations will 
provide stringent bounds on inflationary models based on particle physics
scenarios and probably will validate or rule out specific proposals.

Although the general features of inflation seem robust and universal, the
implementation of the inflationary stage as well as the
departure from scale invariance of the  spectra of scalar and tensor
perturbations depend on the concrete  model and its dynamics. In this 
article we focus on the study of the dynamics of specific
new inflationary scenarios that lead to an inflationary stage after
a phase transition. 

\subsection{Dynamics of Phase Transitions} 
An appealing  new inflation scenario envisages a phase
transition at GUT scales that provides an inflationary
stage\cite{kolb,turner}. The  
concept is an extension of well understood features of phase transitions
in model field theories and has been studied within the context of
inflationary cosmology in a fixed background by Linde and  Vilenkin\cite{linde2,vilenkin} and Guth and Pi\cite{guthpi}. 
Consider a scalar field theory 
(we will neglect fermions, gauge fields and other exotic components in
the discussion) with a typical `Mexican hat' potential that allows
for broken symmetry states. If the system is originally at very
high temperature, larger than critical, i.e. $T_i > T_c$, the
`effective' potential 
has a minimum when the expectation value of the scalar field (the 
order parameter) vanishes. As the temperature cools below critical,
the effective potential develops minima away from the origin and the
expectation value of the scalar field will roll toward the
minima, i.e. the equilibrium values. Consider as a relevant example
the following scalar potential
\begin{eqnarray} 
V_{eff}(\phi,T(t)) & = & \frac{\lambda}{4}\, \phi^4 + 
\frac{1}{2}\, m^2[T(t)]\;\phi^2+ \frac{m^4}{4\lambda} ,
\label{veff} \\
m^2[T]          & = & m^2\left[\frac{T^2}{T^2_c}-1\right],
 \label{masofT} \\
T(t)  & = & \frac{T_i}{a(t)} \; \; ; \; \; T_c \propto \frac{m}{\sqrt{\lambda}},
\end{eqnarray}
where the time dependence in the temperature results from considering
a resummation of the fluctuation contribution to the effective mass in an FRW 
cosmology at finite temperature\cite{guthpi,frwpaper}, akin to the
hard thermal loop resummation. Assuming that the heat bath arises from
particles and radiation at  high temperature and that at temperatures
larger than the critical the system is in local thermodynamic
equilibrium (LTE) with $\phi =0$, Einstein's equation for the Hubble
`constant' becomes 
\begin{equation}
H^2(t) = \frac{8\pi}{3 M^2_{Pl}}\left[g \; T^4(t)+
\frac{m^4_R}{4\lambda} \right],
\end{equation}
with $g$ being a constant that depends on the particle content of the
theory.  At the time at which the phase transition takes place, that is
when $T(t_{pt})= T_c$, the Universe is still dominated by radiation, since
$T^4_c \propto {m^4}/{\lambda^2} >> {m^4}/{\lambda}$ for
weakly coupled theories. Thus we reach the first conclusion: that if
a phase transition is driven by the cooling of the expanding Universe, such
a phase transition will occur during the radiation dominated era. The
inflationary period of exponential (De Sitter) expansion will occur when $T^4(t)<< {m^4}/{4\lambda}$ when the vacuum energy dominates,
which will happen after several e-folds of radiation dominated
expansion, once the effective temperature has red-shifted to almost zero.   

The small inhomogeneities of the CMBR restrict the scalar self
coupling in these models to be $\lambda < 10^{-12}$\cite{kolb} (this will
also be understood later when we compute the amplitude of scalar 
perturbations), and assuming LTE may be unjustified in these weakly
coupled theories. However, even {\em if} LTE holds 
for short wavelength modes of the field theory it is unlikely to
hold for the long wavelength modes. The reason for this is the following:
at the phase transition when $T(t_{pt}) \approx T_c \approx m /
\sqrt{\lambda}$ the Hubble constant is 
$ H \approx m (m /[\lambda M_{Pl}]) $. Assuming the De Sitter stage
of exponential inflation to occur at a GUT scale, this implies that
$m /\lambda^{1/4} \approx 10^{16} \mbox{ Gev }$ with the result that
at the time of the phase transition $d_H(t_{pt}) << m^{-1}$, i.e. 
 the horizon size is much smaller than the Compton wavelength of
the scalar particle. This prevents thermalization of long wavelength
modes even when the short wavelength modes may be strongly coupled to the bath
and  reach LTE. Thus our second conclusion: the phase transition will be 
strongly supercooled, with the long wavelength modes falling quickly
out of LTE, even when LTE prevailed for short wavelength modes. The dynamics
must necessarily be studied away from quasi-equilibrium and any approach
based on effective potentials will miss important physics. 

Thus the phase transition can be described as a sudden quench from
the high temperature phase at $T > T_c$ into the low temperature phase
with $T<< T_c$ on time scales much shorter than the microscopic time
scales for thermalization and relaxation\cite{De Sitter}.
 If the initial state was
disordered in the sense that the expectation value of the order
parameter is zero, then the phase transition occurs with the order
parameter sitting at the top of the potential. Because of the
symmetry $\phi \rightarrow -\phi$ a state with $\phi=\dot{\phi}=0$ will
maintain vanishing value of the order parameter under the dynamics,
and the rolling of the field down its potential hill necessary to
end the inflationary stage must be understood as a consequence of
the quantum fluctuations.

We have argued previously\cite{us1,De Sitter,boylee}
that under these circumstances long wavelength fluctuations will
grow almost exponentially in a manner very similar to spinodal
decomposition and phase separation in condensed matter systems. The
quantum fluctuations as measured by the equal time two-point correlation
function of the field $<\Phi^2(\vec x,t)>$ must 
grow\cite{us1,De Sitter,boylee} in such a way that 
the mean square root fluctuation of the field $\delta \phi =
\sqrt{<\Phi^2(\vec x,t)>}$ will eventually sample the minima of the
potential and reach an equilibrium state. That is to say that at long
times, when 
a quasi-equilibrium state has been achieved, it must be that
$<\Phi^2(\vec x,t)> \approx m^2/ \lambda$. Therefore the quantum
fluctuations must become {\em non-perturbatively} large. Hence we reach
a third conclusion: if the expectation value of the scalar field
is at the top of the potential hill with vanishingly small velocity 
during the inflationary epoch, quantum fluctuations that are responsible
for the process of phase separation will grow non-perturbatively large.
Therefore the dynamics must necessarily be studied within a
non-perturbative framework.  

Although new inflationary scenarios were previously
studied\cite{linde2,vilenkin,stein}, our work is rather different in
that it addresses the description of the dynamics {\em including}
self-consistently  
and non-perturbatively the non-equilibrium growth of quantum
fluctuations and their effect on the dynamics of the scale factor.   

\subsection{The puzzling questions:}

Having reached this conclusion on the non-perturbative growth of the
quantum fluctuations one is presented with  unsettling puzzles: 
typically the scalar field is written as $\Phi(\vec x,t) = \phi(t)+
\delta \phi(\vec x,t)$ where $\phi(t)$ is assumed to be the zero mode
or expectation value of the scalar field and $\delta \phi(\vec x,t)$ the
{\em small} quantum fluctuations that are ultimately responsible for
metric perturbations. Inflation terminates when $\phi(t)$ rolls down 
the potential hill and reaches the minimum, oscillating about it and
eventually decaying into lighter particles leading to the reheating
stage.  

However, the scenario that is envisaged here is that of a quenched or
supercooled phase transition in which $\phi(t) = 0$ throughout
the evolution and the fluctuations grow to be very large and to
eventually sample the minima of the potential. 

 Therefore one is led to the questions: a) what is
truly rolling down?, b) how does inflation end?, c) how sensitive is the
dynamics to the initial value of the expectation value of the scalar
field?, d) if the fluctuations 
grow to become non-perturbatively large will not they provide a large
contribution to the energy momentum tensor and modify the FRW dynamics?,
e) can one make sense of small fluctuations to calculate density
perturbations?. 

Even when the initial value of $\phi \neq 0$, an inflationary stage requires
that it be sufficiently small for the energy density to be dominated by
the vacuum term. Under these circumstances the quantum fluctuations will
nevertheless grow, and the dynamics must address not only
the rolling of the expectation value, but also the growth of spinodal
fluctuations.  

The goal of the present work is to answer all of these
questions. We will see below that it is precisely the large
fluctuations that provide the sensible answers to all of the questions
above, thus reconciling the naive picture of a scalar field rolling
down the hill. Furthermore, a detailed analysis of the dynamics will
allow us to quantify the answer to these questions and, in particular, will
allow a profound interpretation of the effective zero mode, its
initial conditions, and the small fluctuations that will provide the
small density perturbations.   

\section{The Model and Equations of Motion}
Having recognized the non-perturbative dynamics of the long wavelength
fluctuations we need to study the dynamics within a non-perturbative
framework. We require that such framework be: i) renormalizable, ii)
covariant energy conserving, iii) numerically implementable.
There are very few schemes that fulfill all of these criteria: the Hartree
and the large $ N $ approximation\cite{vilenkin,us1,De Sitter}. Whereas the
Hartree approximation is 
basically a Gaussian variational approximation\cite{jackiwetal,guven}
that in general cannot 
be consistently improved upon, the large $ N $ approximation can be
consistently implemented beyond leading order\cite{motola,largen} 
and in our case it has
the added bonus of providing many light fields (associated with
Goldstone modes) that 
will permit the study the effects of other fields which are lighter
than the inflaton on the dynamics. Thus we will study the inflationary
dynamics of a quenched phase transition within the framework of the
large $ N $ limit of a scalar theory in the vector representation of $
O(N) $.

We assume that the universe is spatially flat with a metric given by
eq. (\ref{FRW}).
The matter action and Lagrangian density are given by
\begin{equation}
S_m  =  \int d^4x\; {\cal L}_m = \int d^4x \,
a^3(t)\left[\frac{1}{2}\dot{\vec{\Phi}}^2(x)-\frac{1}{2} 
\frac{(\vec{\nabla}\vec{\Phi}(x))^2}{a^2(t)}-V(\vec{\Phi}(x))\right]
\label{action}
\end{equation}
\begin{equation}
V(\vec{\Phi})  =  \frac{\lambda}{8N}\left(\vec{\Phi}^2+\frac{2N
M^2}{\lambda}\right)^2 
\; \; ; \; \; 
M^2  = - m^2+\xi\;{\cal R}, \label{potential}
\end{equation}
\begin{equation}
{\cal R}  =  6\left(\frac{\ddot{a}(t)}{a(t)}+
\frac{\dot{a}^2(t)}{a^2(t)}\right), \label{ricciscalar}
\end{equation}
where we have included the coupling of $\Phi(x)$ to the scalar curvature ${\cal
R}(t)$ since it will arise as a consequence of renormalization\cite{frwpaper}. 

The gravitational sector includes the usual Einstein term in addition
to a higher order curvature term and a cosmological constant term 
which are necessary to renormalize the theory. The Lagrangian density for
the gravitational sector is therefore:
$$
{\cal L}_g = a^3(t) \left[\frac{{\cal R}(t)}{16\pi G} 
+ \frac{\alpha}{2} {\cal R}^2(t) - K\right].
$$
with K being the cosmological constant.
In principle, we also need to include the terms $R^{\mu\nu}R_{\mu\nu}$
and $R^{\alpha\beta\mu\nu}R_{\alpha\beta\mu\nu}$ as they are also terms
of fourth order in derivatives of the metric (fourth adiabatic order),
but the variations resulting from these terms turn out not to be 
independent of that of ${\cal R}^2$ in the flat
FRW cosmology we are considering.

The variation of the action (\ref{action}) with respect to the 
metric $g_{\mu\nu}$ gives us Einstein's equation
\begin{equation}
\frac{G_{\mu\nu}}{8\pi G} + \alpha H_{\mu\nu} + K g_{\mu\nu}
= - T_{\mu\nu},
\label{extendEinstein}
\end{equation}
where $G_{\mu\nu}$ is the Einstein tensor given by the variation of
$\sqrt{-g}{\cal R}$, $H_{\mu\nu}$ is the higher order curvature term given
by the variation of $\sqrt{-g}{\cal R}^2$, and $T_{\mu\nu}$ is the contribution 
from the matter Lagrangian. 
 With the metric (\ref{FRW}), the various components
of the curvature tensors in terms of the scale factor are:
\begin{eqnarray}
G^{0}_{0} & = & -3(\dot{a}/a)^2, \\
G^{\mu}_{\mu} & = & -{\cal R} = -6\left(\frac{\ddot{a}}{a}
+\frac{\dot{a}^2}{a^2}\right), \\ 
H^{0}_{0} & = & -6\left(\frac{\dot{a}}{a}\dot{{\cal R}} + 
\frac{\dot{a}^2}{a^2}{\cal R} - \frac{1}{12}{\cal R}^2\right), 
\label{hzerozero} \\
H^{\mu}_{\mu} & = & -6\left(\ddot{{\cal R}} + 
3\frac{\dot{a}}{a}\dot{{\cal R}}\right).
\label{htrace}
\end{eqnarray}
Eventually, when we have fully renormalized the theory,
we will set $\alpha_R=0$ and keep as our only contribution to
$K_R$ a piece related to the matter fields which we shall 
incorporate into $T_{\mu\nu}$.  

\subsection{\bf The Large $N$ Approximation} 
To obtain the proper large $N$ limit, the vector field is written as
$$
\vec{\Phi}(\vec x, t) = (\sigma(\vec x,t), \vec{\pi}(\vec x,t)),
$$ 
with $\vec{\pi}$ an $N-1$-plet, and we write
\begin{equation}
\sigma(\vec x,t) = \sqrt{N}\phi(t) + \chi(\vec x,t) \; \; ; \; \; \langle
\sigma(\vec x, t) \rangle= \sqrt{N}\phi(t) \; \; ; \; \; 
\langle \chi(\vec x,t) \rangle = 0.
\label{largenzeromode} 
\end{equation}
To implement the large $N$ limit in a consistent manner, one may introduce an
auxiliary field as in\cite{largen}.  However, the leading order
contribution can be obtained equivalently by invoking the 
factorization\cite{De Sitter,frw2}:

\begin{eqnarray}
\chi^4 & \rightarrow & 6 \langle \chi^2 \rangle \chi^2 +\mbox{ constant },
\label{larg1} \\ \chi^3 & \rightarrow & 3 \langle \chi^2 \rangle \chi,
\label{larg2} \\ \left( \vec{\pi} \cdot \vec{\pi} \right)^2 & \rightarrow &
2 \langle \vec{\pi}^2 \rangle \vec{\pi}^2 - \langle \vec{\pi}^2 \rangle^2+
{\cal{O}}(1/N), \label{larg3} \\ \vec{\pi}^2 \chi^2 & \rightarrow & \langle
\vec{\pi}^2 \rangle \chi^2 +\vec{\pi}^2 \langle \chi^2 \rangle,
\label{larg4} \\ \vec{\pi}^2 \chi & \rightarrow & \langle \vec{\pi}^2
\rangle \chi.  \label{larg5}
\end{eqnarray}

To obtain a large $N$ limit, we define\cite{De Sitter,frw2}
\begin{equation} 
\vec{\pi}(\vec x, t) = \psi(\vec x, t)
\overbrace{\left(1,1,\cdots,1\right)}^{N-1}, \label{filargeN}
\end{equation} 
where the large $N$ limit is implemented by the requirement that
\begin{equation}
\langle \psi^2 \rangle \approx {\cal{O}} (1) \; , \; \langle \chi^2 \rangle
\approx {\cal{O}} (1) \; , \; \phi \approx {\cal{O}} (1).
\label{order1}
\end{equation}
The leading contribution is obtained by neglecting the $ {\cal{O}} ({1}\slash
{N})$ terms in the formal limit. The resulting Lagrangian density is
quadratic, with  linear terms in $\chi$ and $\vec{\pi}$.  
The equations of motion are obtained by imposing the tadpole conditions
$<\chi(\vec x,t)>=0$ and $<\vec{\pi}(\vec x,t)> =0$ which in this case are
tantamount to requiring that the linear terms in $\chi$ and $\vec{\pi}$ in
the Lagrangian density vanish. 
Since the action is quadratic, the quantum fields can be expanded in terms
of creation and annihilation operators and mode functions that obey the
Heisenberg equations of motion
\begin{equation} 
\vec{\pi}(\vec x, t) = \int\frac{d^3k}{(2\pi)^3}
\left[{\vec a}_k f_k(t)e^{i\vec{k}\cdot \vec x} + {\vec a}^{\dagger}_k
f^*_k(t)e^{-i\vec{k}\cdot \vec x} \right] .
\end{equation}
The tadpole condition leads to the following equations
of motion\cite{De Sitter,frw2}:

\begin{equation}
\ddot{\phi}(t)+3H(t)\dot{\phi}(t)+{\cal M}^2(t)\phi(t)=0,
\label{largezeromodeeqn} 
\end{equation}
with the mode functions
\begin{equation}
\left[\frac{d^2}{dt^2}+3H(t)\frac{d}{dt}+\frac{k^2}{a^2(t)}+{\cal
M}^2(t) \right]f_k(t)= 0, 
\label{largenmodes}
\end{equation}
where
\begin{equation}
{\cal M}^2(t) =  -m^2+\xi{\cal R}+ \frac{\lambda}{2}\phi^2(t)+
\frac{\lambda}{2}\langle \psi^2(t) \rangle \; .
\label{Ngranmass}
\end{equation}

In this leading order in $ 1/N $ the theory becomes Gaussian, but with the
self-consistency condition
\begin{equation}
\langle \psi^2(t) \rangle = \int
\frac{d^3k}{(2\pi)^3}\frac{|f_k(t)|^2}{2}.
\label{largenfluc}
\end{equation}

The initial conditions on the modes $f_k(t)$ must now be determined. 
At this stage it proves illuminating to pass to conformal time variables
in terms of the conformally rescaled fields (see \cite{frw2} for a discussion)
in which the mode functions obey an equation which is very similar to that
of harmonic oscillators with time dependent frequencies in Minkowski
space-time.  
It has been realized that different initial conditions on the mode functions
lead to different renormalization counterterms\cite{frw2};
 in particular imposing initial conditions in comoving time leads to
counterterms that depend on these initial conditions. Thus we chose to
impose initial conditions in conformal time in terms of the conformally
rescaled mode functions leading to the following initial conditions
in comoving time:
\begin{equation}
f_k(t_0)=\frac{1}{\sqrt{W_k}}, \;\;\; 
\dot{f}_k(t_0)=\left[-\frac{\dot{a}(t_0)}{a(t_0)}-iW_k\right]f_k(t_0),
\label{initcond}
\end{equation}
with
\begin{equation}
W_k^2 \equiv k^2 + {\cal M}^2(t_0) - \frac{{\cal R}(t_0)}{6}.
\label{freq}
\end{equation}
At this point we recognize that when ${\cal M}^2(t_0) - {\cal R}(t_0)/6 <0$ 
the above initial condition must be modified to avoid imaginary frequencies,
which are the signal of instabilities for long wavelength modes. Thus
we {\em define} the initial frequencies that determine the initial conditions
(\ref{initcond}) as
\begin{eqnarray}
W_k^2 & \equiv &  k^2 + \left|M^2(t_0) - \frac{{\cal
R}(t_0)}{6}\right| \; \mbox{ for } 
k^2 < \left |M^2(t_0) - \frac{{\cal R}(t_0)}{6} \right|\; , \label{unstcond1} \\
W_k^2 & \equiv &  k^2 + M^2(t_0) - \frac{{\cal R}(t_0)}{6} \; \mbox{ for }
k^2 \geq \left |M^2(t_0) - \frac{{\cal R}(t_0)}{6} \right|\; . \label{unstcond2}
\end{eqnarray}

In the large $N$ limit we find the energy density  and pressure 
density to be given by,
\begin{eqnarray}
\frac{\varepsilon}{N} = & & \frac{1}{2}\dot{\phi}^2(t)+
\frac{\lambda}{8}\left(\phi^2(t)+\frac{2M^2}{\lambda}\right)^2  \nonumber \\
+&& \frac{1}{2} \int \frac{d^3k}{2(2\pi)^3 }
\left[ |\dot{f}_k(t)|^2+ \omega^2_k(t)|f_k(t)|^2\right] -
\frac{\lambda}{8}\langle \psi^2(t) \rangle^2,
\label{largenenergy} 
\end{eqnarray}
 
\begin{equation}
\frac{p+\varepsilon}{N} = \dot{\phi}^2(t) + \int \frac{d^3k}{2(2\pi)^3}
 \left[ |\dot{f}_k(t)|^2+
\frac{k^2}{3a^2(t)}|f_k(t)|^2\right]. \label{pplusehart}
\end{equation}
It is straightforward to show that the bare energy is
covariantly conserved by using the equations of motion for the zero mode and
the mode functions. 

\section{Renormalization}

Renormalization is a very subtle but important issue in gravitational
backgrounds\cite{birriel}. The fluctuation contribution
$\langle \psi^2(\vec x,t) \rangle$, the energy, and the pressure all need to be
renormalized. The renormalization aspects in curved space times have been
discussed at length in the literature\cite{birriel} and has
been extended to the large $N$ self-consistent approximations for
the non-equilibrium backreaction problem in\cite{largen,frw2}.
More recently a consistent and covariant regularization scheme that
can be implemented numerically has been provided
(see J. Baacke's contribution to these proceedings and\cite{baacke,ramsey}).

In terms of the effective mass term for the large $ N $ limit given by
(\ref{Ngranmass}) and defining the quantity
\begin{eqnarray}
B(t) &\equiv& a^2(t)\left({\cal M}^2(t)-{\cal{R}}/6 \right),\label{boft}\\
{\cal M}^2(t) &=& -m^2_B+\xi_B {\cal R}(t)+\frac{\lambda_B}{2}\phi^2(t)
+\frac{\lambda_B}{2}\langle\psi^2(t)\rangle_B \; , \label{masso}
\end{eqnarray}
where the subscript $B$ stands for bare quantities,  
we find the following large $k$ behavior for the case of an {\em arbitrary}
scale factor $a(t)$ (with $a(0)=1$):
\begin{eqnarray}
|f_k(t)|^2 &=& \frac{1}{ka^2(t)}+ \frac{1}{2k^3
a^2(t)}\left[-B(t) \right] \cr \cr &+&
{1 \over {8 a(t)^2 \; k^5 }}\left\{ B(t)[ 3 B(t)  ] + a(t)
\frac{d}{dt} \left[ a(t) {\dot B}(t) \right]  \right\} + {\cal{O}}(1/k^7)
\label{sub1}\\ 
|\dot{f}_k(t)|^2 &=&
\frac{k}{a^4(t)}+\frac{1}{ka^2(t)}\left[H^2(t)+
\frac{1}{2}\left({\cal M}^2(t)-{\cal{R}}/6 \right) \right] \cr \cr & + & {1 \over {8 a(t)^4 \; k^3 }}\left\{ - B(t)^2 - a(t)^2 {\ddot B}(t) + 3 a(t)
{\dot a}(t)
{\dot B}(t) - 4 {\dot a}^2(t) B(t) \right\} +  {\cal{O}}(1/k^5).  \label{sub2}
\end{eqnarray}

Although the divergences can be dealt with by dimensional
regularization, this procedure is not well suited to numerical
analysis (see however ref.\cite{baacke}).  We will make our 
subtractions using an ultraviolet cutoff constant in {\em physical
coordinates}. This guarantees that the counterterms will be time
independent. The renormalization then proceeds much in the same manner as in
reference\cite{frwpaper}; the quadratic divergences renormalize the
mass and the 
logarithmic terms renormalize coupling constant and the coupling to the Ricci
scalar. The renormalization conditions on the mass, coupling to the Ricci
scalar and coupling constant are obtained from the requirement that the
frequencies that appear in the mode equations are finite\cite{frwpaper}, i.e:
\begin{equation}
-m^2_B+\xi_B {\cal R}(t)+\frac{ \lambda_B}{2}\phi^2(t)
+\frac{\lambda_B}{2}\langle\psi^2(t)\rangle_B=
-m^2_R+\xi_R {\cal R}(t)+\frac{ \lambda_R}{2}\phi^2(t)
+\frac{\lambda_R}{2}\langle\psi^2(t)\rangle_R \label{rencond}
\end{equation}

Finally we arrive at the following set of renormalizations\cite{frw2}:
\begin{eqnarray}
\frac{1}{8\pi G_R} &=& \frac{1}{8\pi G_B} 
- 2\left(\xi_R-\frac16\right)\frac{\Lambda^2}{16\pi^2}
- 2\left(\xi_R-\frac16\right)m_R^2\frac{\ln(\Lambda/\kappa)}{16\pi^2}, \\
\alpha_R &=& \alpha_B 
- \left(\xi_R-\frac16\right)^2\frac{\ln(\Lambda/\kappa)}{16\pi^2}, \\
K_R &=& K_B - \frac{\Lambda^4}{16\pi^2}
- m_R^2\frac{\Lambda^2}{16\pi^2}
+ \frac{m_R^4}{2}\frac{\ln(\Lambda/\kappa)}{16\pi^2}, \\
m_R^2 &=& m_B^2 + \lambda_R\frac{\Lambda^2}{16\pi^2}
- \lambda_R m_R^2\frac{\ln(\Lambda/\kappa)}{16\pi^2}, \\
\xi_R &=& \xi_B 
- \lambda_R\left(\xi_R-\frac16\right)\frac{\ln(\Lambda/\kappa)}{16\pi^2}, \\
\lambda_R &=& \lambda_B - \lambda_R\frac{\ln(\Lambda/\kappa)}{16\pi^2},\\
\langle \psi^2(t) \rangle &=& \int
\frac{d^3k}{2(2\pi)^3}\left\{|f_k(t)|^2-\left[
\frac{1}{ka^2(t)}- \frac{\Theta(|\vec k|-\kappa)}{2k^3a^2(t)}
\left[{\cal M}^2(t)-{\cal{R}}/6 \right]
\right] \right\} .
\end{eqnarray}
Here, $\kappa$ is the renormalization point.  As expected, the logarithmic 
terms are consistent with the renormalizations found using dimensional
regularization\cite{baacke,ramsey}.  Again, we set $\alpha_R=0$ and
choose the renormalized cosmological constant such that the vacuum
energy is zero in the true vacuum.  We emphasize that while 
the regulator we have chosen does not respect the covariance of 
the theory, the renormalized energy momentum tensor defined in this 
way nevertheless retains the property of covariant conservation in the
limit when the cutoff is taken to infinity.
In what follows, we drop the subscripts from the renormalized
parameters.

 The logarithmic subtractions can be neglected because of the coupling
$\lambda \leq 10^{-12}$.  Using the Planck scale as the cutoff and the
inflaton mass $m_R$ as a renormalization point, these terms are of order
$\lambda \ln[M_{pl}/m_R] \leq 10^{-10}$, for $m \geq 10^9 \mbox{ GeV }$. An
equivalent statement is that for these values of the coupling and inflaton
masses, the Landau pole is well beyond the physical cutoff $M_{pl}$.
Our relative 
error in the numerical analysis is of order $10^{-8}$, therefore our numerical
study is insensitive to the logarithmic corrections. Though these corrections
are fundamentally important, numerically they can be neglected. Therefore, in
what follows, we will neglect logarithmic renormalization and subtract only
quartic and quadratic divergences in the energy and pressure, and quadratic
divergences in the fluctuation contribution.

\section{Renormalized Equations of Motion for Dynamical Evolution}
It is convenient to introduce the following dimensionless quantities
and definitions,
\begin{equation}
\tau = m_R t \quad ; \quad h= \frac{H}{m_R} \quad ; 
\quad q=\frac{k}{m_R} \quad \; \quad
\omega_q = \frac{W_k}{m_R} \quad ; \quad g= \frac{\lambda_R}{8\pi^2} \; ,
\label{dimvars1}
\end{equation}
\begin{equation}
\eta^2(\tau) = \frac{\lambda_R}{2m^2_R} \; \phi^2(t)
\quad ; \quad  g\Sigma(\tau) = \frac{\lambda}{2m^2_R}\; \langle \psi^2(t)
\rangle_R  \quad ; \quad f_q(\tau) \equiv \sqrt{m_R} \; f_k(t) \; .
\label{dimvars3}
\end{equation}

Choosing $\xi_R=0$ (minimal coupling)  and the renormalization
 point $\kappa = |m_R|$ and setting $a(0)=1$, 
the equations of motion become:

\vspace{2mm} 

\begin{equation}
\left[\frac{d^2}{d \tau^2}+ 3h \frac{d}{d\tau}-1+\eta^2(\tau)+
g\Sigma(\tau)\right]\eta(\tau) = 0, 
\label{zeromode}
\end{equation}

\begin{eqnarray} 
& & \left[\frac{d^2}{d \tau^2}+3h
\frac{d}{d\tau}+\frac{q^2}{a^2(\tau)}-1+\eta^2+g\Sigma(\tau)
\right]f_q(\tau)=0, \nonumber \\ 
& &  f_q(0)  =  \frac{1}{\sqrt{\omega_q}} \quad ; \quad 
\dot{f}_q(0)  = \left[-h(0)-i\omega_q\right]f_q(0), \nonumber \\
& & \omega_q  =  
\left[q^2-1+\eta^2(0)-\frac{{\cal
R}(0)}{6m^2_R}+g\Sigma(0)\right]^{\frac{1}{2}} \; \mbox{ for } \; q^2
> -1+\eta^2(0)-\frac{{\cal R}(0)}{6m^2_R}+g\Sigma(0), \nonumber \\ 
& & \omega_q  =  
\left[q^2+1-\eta^2(0)+\frac{{\cal
R}(0)}{6m^2_R}-g\Sigma(0)\right]^{\frac{1}{2}} \; \mbox{ for } \; q^2
< -1+\eta^2(0)-\frac{{\cal R}(0)}{6m^2_R}+g\Sigma(0) . 
\label{modes}  
\end{eqnarray}
The initial conditions for $\eta(\tau)$ will be specified later. 
An important point to notice is that the equation of
motion for the $q=0$ mode coincides with that of the zero mode
(\ref{zeromode}). Furthermore, for $\eta(\tau \rightarrow \infty) \neq
0$, a stationary (equilibrium) solution of the eq. (\ref{zeromode})  
is obtained when the sum rule\cite{us1,De Sitter,frw2}
\begin{equation}
-1+\eta^2(\infty)+g\Sigma(\infty) = 0 \label{sumrule}
\end{equation}
is fulfilled. This sum rule is nothing but a consequence of Goldstone's
theorem and is a result of the fact that the large $ N $ approximation 
satisfies the Ward identities associated with the $ O(N) $ symmetry, since
the term  $-1+\eta^2+g\Sigma$ is seen to be the effective mass of the
modes transverse to the symmetry breaking direction, i.e. the Goldstone
modes in the broken symmetry phase.

In terms
of the zero mode $\eta(\tau)$ and the quantum mode function given
by eq.(\ref{modes}) we find that the Friedmann equation for the dynamics
of the scale factor in dimensionless variables is given by

\begin{equation}
 h^2(\tau)    =   4h^2_0 \epsilon_R(\tau) \quad  ;   \quad h^2_0 =
 \frac{4\pi N m^2_R}{3M^2_{Pl}\lambda_R} \label{hubblequation} 
 \end{equation}
and the renormalized energy and pressure are given by: 

\begin{eqnarray}
 \epsilon_R(\tau) &  =  &  
\frac{1}{2}\dot{\eta}^2+\frac{1}{4}\left(-1+\eta^2+g\Sigma \right)^2+
 \nonumber \\ 
& &  \frac{g}{2}\int q^2 dq \left[\left(|\dot{f_q}|^2-
{\cal S}^{(1)}(q,\tau)\right)
+\frac{q^2}{a^2}\left(|f_q|^2-{\cal S}^{(2)}(q,\tau)\right) \right]\; ,
\label{hubble} \\
 (p+\varepsilon)_R  & = & \frac{2Nm^4_R}{\lambda_R}\left\{
\frac{1}{2}\dot{\eta}^2+ g \int q^2 dq \left[|\dot{f_q}|^2-
{\cal S}^{1}(q,\tau)
+\frac{q^2}{3a^2}\left(|f_q|^2-{\cal S}^{(2)}(q,\tau)\right)
\right]\right\}, \label{ppluse} 
\end{eqnarray}
where the subtractions ${\cal S}^{(1)}$ and ${\cal S}^{(2)}$ are given
by the right hand sides of eqns.(\ref{sub2}) and (\ref{sub1}) respectively.
 
We want to emphasize the following aspects of the set of equations
(\ref{zeromode},\ref{modes},\ref{hubble}): they
determine the full dynamics of the matter plus  classical gravity 
including backreaction effects both on  the metric as well as in the
dynamics of the fields, they are fully
renormalized, maintain covariant conservation and in principle they can be
consistently improved in the $ 1/N $ expansion. 

In order to provide the full
solution we now must provide the values of $\eta(0)$, $\dot{\eta}(0)$,
and $h_0$. Assuming that the 
inflationary epoch is associated with a phase transition at the GUT scale,
this requires that $ N m^4_R/\lambda_R \approx 
(10^{15}\mbox{ Gev })^4 $ and assuming the bound on the scalar
self-coupling $\lambda_R \approx 10^{-12}-10^{-14}$ (this will be seen
later 
to be a compatible requirement), we find that $h_0 \approx N^{1/4}$ which
we will take to be reasonably given by $h_0 \approx 1-10$ (for example
in popular GUT's $ N \approx 20 $ depending on particular representations).

We will begin by studying the case of most interest from the point of view
of describing the phase transition: $\eta(0)=0$ and $\dot{\eta}(0)=0$,
which are the initial conditions that led to our puzzling questions. With
these initial conditions, the evolution equation for the zero mode
eq. (\ref{zeromode}) determines that $\eta(\tau) = 0$ by symmetry.

\subsection{Early time dynamics:}
Before engaging in a full numerical study, it proves illuminating to
obtain an estimate of the relevant time scales and an intuitive idea of
the main features of the dynamics. Because the coupling is so weak and after
renormalization the contribution from the quantum fluctuations to 
the equations of motion 
is finite, we can obtain an estimate of the early time dynamics by
neglecting the backreaction terms in the equations for the mode
functions (\ref{modes}) and the Hubble constant (\ref{hubblequation}). 
Setting $\eta = 0$ and
$g\Sigma \approx 0$ in eq. (\ref{modes}) and also setting to zero the
terms 
proportional to $g$ in eq.(\ref{hubble}), the evolution equations for the
mode functions are those for an inverted oscillator in De Sitter space-time,
which have been studied by Guth and Pi\cite{guthpi}. One
obtains the approximate solution 
\begin{eqnarray}
h(t) & \approx & h_0, \nonumber \\
f_q(t) & \approx & e^{-3 h_0\tau/2} \left[A_q \;
J_{\nu}\left(\frac{q}{h_0}e^{-h_0\tau}\right)+ 
B_q  \; J_{-\nu}\left(\frac{q}{h_0}e^{-h_0\tau}\right)\right], \nonumber \\
\nu & = & \sqrt{\frac{9}{4}+\frac{1}{h^2_0}}, \label{earlytime} 
\end{eqnarray}
where $ J_{\pm \nu}(z) $ are Bessel functions, and $A_q$ and $B_q$ are
determined by the initial conditions on the mode functions.

When the physical wavevectors cross the horizon, i.e. when $qe^{-h_0
\tau}/h_0 \ll 1$ we find that the mode functions factorize: 
\begin{equation}
f_q(\tau) \approx  {{B_q} \over {\Gamma(1-\nu)}} 
\; \left( {{2h_0\, }\over q}\right)^{\nu}e^{(\nu-3/2)h_0 \tau}. \label{factor}
\end{equation}
 This  result reveals a very
important feature: $\nu > 3/2$; because of the negative squared mass
term in the matter Lagrangian leading to symmetry breaking, we see
that all of the mode functions {\em grow exponentially} after horizon
crossing (for positive squared mass they would {\em decrease
exponentially} after horizon crossing). This exponential growth is a 
consequence of the spinodal instabilities which is modified in De Sitter
space-time but is a hallmark of the process of phase separation that
occurs to complete the phase transition. 
 We note, in addition that the time 
dependence is exactly given by that of the $ q=0 $ mode, i.e. the zero 
mode, which is a consequence of the redshifting of the wavevectors and 
the fact that after horizon crossing the contribution of the term
$q^2/a^2(\tau)$ in the equations of motion become negligible.
 Then we clearly  see that the quantum fluctuations grow exponentially and
they will begin to be of the order of the tree level terms in the
equations of motion when $g\Sigma(\tau) \approx 1$. At large
times $\Sigma(\tau) \approx {\cal F}^2(h_0) (h_0/2\pi)^2 e^{(2\nu-3)h_0 \tau}$, with
${\cal F}(h_0)$ a finite constant that depends on the initial conditions and
is found numerically to be of ${\cal O}(1)$.  

In terms of the initial dimensionfull variables, this condition translates
to $<\psi^2(\vec x,t)>_R \approx m^2_R/\lambda_R$, i.e. the quantum
fluctuations sample the minima of the (renormalized) tree level potential.
We find that the
time at which the contribution of the 
quantum fluctuations becomes of the same order as the tree level terms is
estimated to be\cite{De Sitter}
\begin{equation}
\tau_s \approx \frac{1}{(2\nu-3)h_0}
\ln\left(\frac{32\pi^4}{\lambda h_0^2 {\cal F}^2(h_0)}\right) 
= \frac32 h_0 \ln\left(\frac{32\pi^4}{\lambda h_0^2 {\cal F}^2(h_0)}\right) 
+ {\cal O}(1/h_0).
\label{spinodaltime}
\end{equation}
At this time, the contribution of the quantum fluctuations makes the
back reaction very important and, as will be seen numerically, this
translates into the fact that $\tau_s$ also determines the end of the
De Sitter era and the end of inflation. The total number of e-folds during
the stage of exponential expansion of the scale factor (constant
$h_0$) is  given by   
\begin{equation}
N_e \approx \frac{1}{(2\nu-3)}
\ln\left(\frac{32\pi^4}{\lambda h_0^2 {\cal F}^2(h_0)}\right) 
= \frac32 h_0^2 \ln\left(\frac{32\pi^4}{\lambda h_0^2 {\cal F}^2(h_0)}\right) 
+ {\cal O}(1/h_0)\label{efolds}
\end{equation}
For large $h_0$ we see that the number of efolds scales as $h^2_0$ as well
as with the logarithm of the inverse coupling. 
These results (\ref{factor},\ref{spinodaltime},\ref{efolds}) will be
confirmed numerically below and will be of paramount importance for the
interpretation of the main consequences of the dynamical evolution. 

\subsection{$\eta(0) \neq 0$: classical or quantum behavior?}

Above we have analyzed the situation when $\eta(0) =0$ (or in dimensionfull
variables $\phi(0)=0$). The typical analysis of inflaton dynamics in
the literature involves the {\em classical} evolution of $\phi(t)$ 
with an initial condition in which $\phi(0)$ is very close to zero (i.e.
the top of the potential hill) in the `slow-roll' regime, for which
$ \ddot{\phi} \ll 3H\dot{\phi}$. Thus, it is important
to quantify the initial conditions on $\phi$ for which the dynamics will be
determined by the classical evolution of $\phi$ and those for which the quantum
fluctuations dominate the dynamics. We can provide a criterion to
separate classical from quantum dynamics by analyzing the relevant time
scales, estimated by neglecting
non-linearities and backreaction effects. Considering the linear
evolution 
of the zero mode in terms of dimensionless variables, and considering
$\eta(0) \neq 0$ ($\dot{\eta}(0) \neq 0$ simply corresponds
to a shift in origin of time), we find
\begin{equation}
\eta(\tau) \approx \eta(0) e^{(\nu - \frac{3}{2})h_0\tau}.
\end{equation}
The non-linearities will become important and eventually terminate
inflation when $\eta(\tau) \approx 1$. This corresponds to a time
\begin{equation}
\tau_c \approx \frac{\ln(1/ \eta(0))}{(\nu - \frac{3}{2})h_0}.
\label{classtime}
\end{equation}
 If $ \tau_c $ is much smaller than the spinodal time $ \tau_s $ given
by eq.(\ref{spinodaltime}) then the {\em classical} evolution of the
zero mode will dominate the dynamics and the quantum fluctuations will
not 
become very large, although they will still undergo spinodal growth. On 
 the other hand, if $\tau_c \gg \tau_s$ the quantum fluctuations will
grow to be very large well before the zero mode reaches the non-linear
regime. In this case the dynamics will be determined completely by
the quantum fluctuations. Then the criterion for the classical or quantum
dynamics is given by
\begin{eqnarray} 
\eta(0) & \gg & \sqrt{\lambda}h_0 \Longrightarrow \mbox{ classical dynamics }
\nonumber \\
\eta(0) & \ll & \sqrt{\lambda}h_0 \Longrightarrow \mbox{ quantum dynamics }
\label{classquandyn} 
\end{eqnarray}
or in terms of dimensionfull variables $\phi(0) \gg H_0$ leads to 
{\em classical dynamics} and $\phi(0) \ll H_0$ leads to 
{\em quantum dynamics}. 

However, even when the classical evolution of the
zero mode dominates the dynamics, the quantum fluctuations grow
exponentially after horizon crossing unless the value of $\phi$ is
very close to the minimum of the tree level potential. In the large $
N $ approximation the spinodal line, that is the values of $\phi$ for  
which there are spinodal instabilities, reaches all the way to the minimum
of the tree level potential as can be seen from the equations of motion for
the mode functions. 
Therefore even in the
classical case one must understand how to deal with quantum fluctuations
that grow and become large after horizon crossing.  

\subsection{Numerics}
The time evolution is carried out by means of a fourth order Runge-Kutta
routine with adaptive stepsizing while the momentum 
integrals are carried out using an 11-point Newton-Cotes integrator.  
The relative errors in both
the differential equation and the integration are of order $10^{-8}$.
We find that the energy is covariantly conserved throughout the evolution
to better than a part in a thousand. Figures (1-3) show $g\Sigma(\tau)$ 
vs $\tau$,
$h(\tau)$ vs. $\tau$ and $\ln(|f_q(\tau)|^2)$ vs $\tau$ for several values
of $q$ with larger $q's$ corresponding to successively lower curves. 
Figures (4,5) show $p(\tau)/\varepsilon(\tau)$ and the horizon size 
$h^{-1}(\tau)$ for 
$\lambda = 10^{-12} \; ; \; \eta(0)=0 \; ; \; \dot{\eta}(0)=0$
and we have chosen the representative value $h_0=1.5$.

Figures 1 and 2 show clearly that when the contribution of the quantum
fluctuations $g\Sigma(\tau)$ becomes of order 1 inflation basically ends,
and the time scale for $g\Sigma$ to reach ${\cal O}(1)$ is very well
described by  the estimate (\ref{spinodaltime}). From figure 1 we see
that this happens for $\tau \approx 75$, leading to a number of
e-folds  
$N_e \approx 110$ which is correctly estimated by  (\ref{spinodaltime},
\ref{efolds}). 

Figure 3 shows clearly the factorization of the modes after they
cross the horizon given by eq. (\ref{factor}).
 The slopes of all the curves after they become
straight lines in figure 3 is given exactly by $(2\nu-3)$, whereas the
intercept depends on the initial condition on the mode function and
the larger the value of $ q $ the smaller the intercept because the
amplitude of the mode function is smaller initially. Although the
intercept depends on the initial conditions on the long-wavelength
modes, the slope is independent of the value of $q$ and is the same as
what would be obtained in the linear approximation for the {\em
square} of the zero mode at times 
long enough that the decaying solution can be neglected but short enough
that the effect of the non-linearities is very small.
 Notice from the figure that when inflation ends and
the non-linearities become important all of the modes basically saturate.
This is also what one would expect from the solution of the zero mode:
exponential growth in early-intermediate times (neglecting the
decaying solution), with a growth exponent
given by $(\nu - 3/2)$ and an asymptotic behavior of small oscillations
around the equilibrium position, which for the zero mode is $\eta =1$, but
for the $q \neq 0$ modes depends on the initial conditions. 
All of the mode functions have this behavior once they cross the horizon.
We have also studied the phases of the mode functions and we found that 
they `freeze' after horizon crossing. This is natural since both the
real and imaginary parts of $f_q$ obey the same equation but with different
boundary conditions. After the physical wavelength crosses the horizon, the
dynamics is insensitive to the value of $q$ for real and imaginary parts and
the phases become independent of time. Again, this is a consequence of
factorization.  

The growth of the quantum fluctuations is sufficient to end inflation
at a time given by $\tau_s$ in eq. (\ref{spinodaltime}). Furthermore figure
4 shows that during the inflationary epoch $p(\tau)/\varepsilon(\tau) 
\approx -1$ and the end of inflation is rather sharp at $\tau_s$ with
$p(\tau)/\varepsilon(\tau)$ oscillating between $\pm 1$ with zero average
over the cycles, resulting in matter domination. Figure 5 shows this
feature very clearly; $h(\tau)$ is constant during the De Sitter epoch and
becomes matter dominated after the end of inflation with $h^{-1}(\tau) 
\approx 3(\tau -\tau_f) / 2$, with $\tau_f$ the time at the end of inflation. There are small oscillations around this value because both $p(\tau)$ and $\varepsilon(\tau)$ oscillate. These oscillations
are a result of small oscillations of the mode functions after they
saturate, and are also a
feature of the solution for a zero mode. 

\subsection{Zero Mode Assembly}
This remarkable feature of factorization of the mode functions after
horizon crossing can be elegantly summarized as
\begin{equation}
f_k(t)|_{k_{ph}(t) \ll H} = g(q,h)f_0(\tau),\label{factor2}
\end{equation}
with $f_0(\tau)$ a real function that obeys the zero mode equation
(\ref{zeromode}), $k_{ph}(t) = ke^{-Ht}$  the physical momentum, 
and $ g(q,h)$ a complex constant.
 Since the factor $g(q,h)$ depends solely on the initial
conditions on the mode functions, it turns out that for two mode
functions corresponding to momenta $k_1,k_2$ that have crossed the
horizon at times $t_1 > t_2$, the ratio of the two mode functions  at
time $t_s>t >t_1 >t_2$ is $f_{k_1}(t)/f_{k_2}(t) \propto e^{(\nu-\frac{3}{2})h (\tau_1 - \tau_2)} > 1$. Then if we consider the
contribution of these modes to the  {\em renormalized} quantum fluctuations a long time after the beginning of inflation (so as to neglect the decaying solutions), we find that $g\Sigma(\tau) \approx {\cal C}e^{(2\nu-3)h \tau} + \mbox{ small }$, where `small' stands for
the contribution of mode functions associated with momenta that have not
yet crossed the horizon at time $\tau$, which give a perturbatively
small (of order $\lambda$) contribution.  Then it is clear that after
several e-folds from the beginning of inflation, we can define an
`effective zero mode' as 
\begin{equation}
\eta^2_{eff}(\tau) \equiv g\Sigma(\tau), \mbox{ or in dimensionfull
variables, } \phi_{eff}(t) \equiv \left[\langle \psi^2(\vec x, t) \rangle_R \right]^{\frac{1}{2}}.
\label{effectivezeromode}
\end{equation}
Although this identification seems natural, we emphasize that it
is by no means a trivial or ad-hoc statement. There are several
important features that allow an {\em unambiguous} identification:
i) $\left[\langle \psi^2(\vec x, t) \rangle_R \right]$ is a fully 
renormalized operator and hence finite, ii) because of  the factorization
 of the superhorizon modes that enter in the evaluation of $\left[\langle \psi^2(\vec x, t) \rangle_R \right]$,  $\phi_{eff}(t)$ (\ref{effectivezeromode}) 
{\em obeys the equation of motion for the zero mode}, iii) this identification is valid several e-folds after the beginning of inflation,
after the transient decaying solutions have died away and the integral
in $\langle \psi^2(\vec x,t) \rangle$
is dominated by the modes with wavevector $k$ that have crossed the horizon at $t(k) << t$.
Numerically we see that this identification holds throughout the
dynamics but for a very few e-folds at the beginning of inflation. This
factorization determines at once the initial conditions of the effective
zero mode that can be extracted numerically: after the first few efolds and
long before the end of inflation we find
\begin{equation}
 \phi_{eff}(t) \equiv \phi_{eff}(0)e^{(\nu-
\frac{3}{2})Ht} \; \; ; \; \; \phi_{eff}(0) \equiv \frac{H}{2\pi} {\cal F}(H/m),
\label{effzeromodein}
\end{equation}
where we parametrized $\phi_{eff}(0) \equiv \frac{H}{2\pi} {\cal F}(H/m)$ to make contact with the literature.
 We find numerically that ${\cal F}(H/m)  \approx
{\cal O}(1)$ for a large range of $0.1 \leq H/m \leq 50$ and depends on the 
initial conditions of the long wavelength modes. 

 The factorization of the superhorizon modes (that dominate the integral) implies that
\begin{eqnarray}
g\int q^2 dq |f^2_q(\tau)| & \rightarrow &  C^2_0 \; |f^2_0(\tau)|,
\label{int1} \\  
g\int q^2 dq |\dot{f}^2_q(\tau)| & \rightarrow & C^2_0\; |\dot{f}^2_0(\tau)|,
\label{int2} \\
g\int \frac{q^4}{a^2(\tau)} dq |f^2_q(\tau)| & \rightarrow &  C^2_1\;
\frac{|f^2_0(\tau)|}{a^2(\tau)}. \label{int3} 
\end{eqnarray}
We also find numerically that even when $\eta(0) \neq 0$ this 
factorization phenomenon is very robust,
and after a few e-folds from the beginning of inflation the dynamics is
completely determined by the effective zero mode
\begin{equation}
\eta_{eff}(\tau) \equiv \sqrt{\eta^2(\tau)+g\Sigma(\tau)}.
\end{equation}
We have checked numerically that the dynamics  of the scale factor and
equation of state obtained from the full
quantum problem is exactly equivalent to that obtained from the classical
problem in terms of $\eta_{eff}(\tau)$. We have also checked numerically
that the estimate for the classical to quantum crossover given by
eq.(\ref{classquandyn}) is quantitatively correct. Thus in the
classical case in 
which $\eta(0) \gg \sqrt{\lambda}$ we find that $\eta_{eff}(\tau) =
\eta(\tau)$ whereas in the opposite, quantum case $\eta_{eff}(\tau) = 
\sqrt{g\Sigma(\tau)}$. We have run the numerical evolution of the scale
factor with only the {\em classical} equation for $\eta_{eff}$ with the
proper initial conditions and found that it coincides within our numerical
error with the evolution obtained by the full system of equations in either
the classical or quantum case. 

This remarkable feature of the zero mode assembly of long-wavelength
spinodally unstable modes is a consequence of the presence of the horizon.
It also explains why despite the fact that asymptotically when the
fluctuations sample the broken symmetry state, the equation of state is
that of matter. 
Since the excitations in the broken symmetry state are massless Goldstone
bosons one would expect radiation domination. However, the `assembly'
phenomenon, i.e. the redshifting of the wave vectors, makes these modes behave
exactly like zero momentum modes that give an equation of state of matter
domination (upon averaging over the small oscillations around the minimum).  

\section{Making sense of `small fluctuations':}

Having recognized the effective classical variable that can be interpreted
as the component of the field that drives the FRW background and rolls
down the classical potential hill, we want to recognize unambiguously
the small fluctuations. We have argued above that after horizon crossing,
all of the mode functions evolve proportionally to the zero mode,
and the question arises: which modes are assembled into the effective
zero mode and which modes are treated as perturbations? In principle every
$k\neq 0$ mode provides some spatial inhomogeneity, and assembling these
into an effective homogeneous zero mode seems in principle to do away with
the very inhomogeneities that one wants to study. However, scales of cosmological importance today have first crossed the horizon during the
last 60 or so e-folds of inflation. Recently Grishchuk\cite{grishchuk}
 has argued that the
sensitivity of the measurements of $\Delta T/T$ probe inhomogeneities on
scales $\approx 500$ times the size of the present horizon. Therefore scales
that are larger than these and that have first crossed the horizon much earlier than the last 60 e-folds of inflation are unobservable today and
can be treated as an effective homogeneous component, whereas the scales that
can be probed experimentally via the CMB inhomogeneities today must be treated 
separately as part of the inhomogeneous perturbations of the CMB. 

Thus a consistent description of the dynamics in terms of an effective
zero mode plus `small' quantum fluctuations can be given provided:
a) the total number of e-folds $N_e \gg 60$, b) all the modes that have
crossed the horizon {\em before} the last 60-65 e-folds are assembled into
an effective {\em classical} zero mode via $\phi_{eff}(t) = 
\left[\phi^2_0(t)+ \langle \psi^2(\vec x,t) \rangle_R
\right]^{\frac{1}{2}}$, c) the modes that cross the horizon during the
last 60--65 e-folds are accounted as `small' perturbations. The reason
for the requirement a) is that in the separation 
$\phi(\vec x, t) = \phi_{eff}(t)+\delta \phi(\vec x,t)$ one requires that
$\delta \phi(\vec x,t)/\phi_{eff}(t) \ll 1$. As argued above after the 
modes cross the horizon, the ratio of amplitudes of the mode functions remains
constant and given by $e^{(\nu - \frac{3}{2})\Delta N}$ with $\Delta N$ 
being the number of e-folds between the crossing of the smaller $ k $ and the
crossing of the larger $ k $. Then for $\delta \phi(\vec x, t)$ to be much
smaller than the effective zero mode, it must be that the Fourier components
of $\delta \phi$ correspond to very large $k$'s at the beginning of inflation,
so that the effective zero mode can grow for a long time before the components
of $\delta \phi$ begin to grow under the spinodal instabilities. 
In fact requirement a) is not very severe; in the figures (1-5) we have taken
$h_0 = 1.5$ which is a very moderate value and yet for $\lambda = 10^{-12}$
the inflationary stage lasts for over 100 e-folds, and as argued above, the
larger $h_0$ for fixed $\lambda$, the longer is the inflationary stage. 
Therefore under this set of conditions, the classical dynamics of the effective zero mode $\phi_{eff}(t)$  drives the FRW background, whereas
the inhomogeneous fluctuations $\delta \phi(\vec x,t)$, which are made up
of Fourier components with wavelengths that are much smaller than the
horizon at the beginning of inflation and that cross the horizon during
the last 60 e-folds, provide the inhomogeneities that seed density
perturbations.   

\section{Scalar Metric Perturbations:}
Having identified the effective zero mode and the `small perturbations',
we are now in position to provide an estimate for the amplitude and spectrum
of scalar metric perturbations. We use the clear formulation by Mukhanov,
Feldman and Brandenberger\cite{mukhanov} in terms of gauge invariant
variables. In particular we focus on the dynamics of the Bardeen
potential\cite{bardeen}, which in longitudinal gauge is identified
with the 
Newtonian potential. The equation of motion for the Fourier components (in
terms of comoving wavevectors) for this variable in terms of the effective
zero mode is\cite{mukhanov}
\begin{equation}
\ddot{\Phi}_k +
\left[H(t)-2\frac{\ddot{\phi}_{eff}(t)}{\dot{\phi}_{eff}(t)}\right] 
\dot{\Phi}_k+\left[\frac{k^2}{a^2(t)}+ 2\left(\dot{H}(t)-H(t)
\frac{\ddot{\phi}_{eff}(t)}{\dot{\phi}_{eff}(t)}\right)\right]\Phi_k =
0 .
\label{bardeen}
\end{equation}

We are interested in determining the dynamics of $\Phi_k$ for those
wavevectors that cross the horizon during the last 60 e-folds before the
end of inflation. During the inflationary stage the numerical analysis
suggests that to a very good approximation
\begin{equation}
H(t) \approx H_0  \; ; \; \phi_{eff}(t) = \phi_{eff}(0)e^{(\nu-
\frac{3}{2})H_0t}, \label{infla}
\end{equation}
where $H_0$ is the value of the Hubble constant during inflation, leading to 
\begin{equation}
\Phi_k(t) = e^{(\nu -2)H_0t}\left[a_k
H^{(1)}_{\beta}\left(\frac{ke^{-H_0t}}{H_0}\right) 
+b_k H^{(2)}_{\beta}\left(\frac{ke^{-H_0t}}{H_0}\right)\right]
\; ; \; \beta= \nu-1 \; .
\label{solbardeen}
\end{equation}
The coefficients $a_k,b_k$ are determined by the initial conditions.

Since we are interested in the wavevectors that cross the horizon during
the last 60 e-folds, the consistency for the zero mode assembly and
the interpretation of `small perturbations' requires that there must be
many e-folds before the {\em last} 60. We are then considering wavevectors
that were deep inside the horizon at the onset of inflation. 
 Mukhanov et. al.\cite{mukhanov} show that $\Phi_k(t)$ is related to
the canonical `velocity field' that determines scalar  perturbations
of the metric and which is quantized with Bunch-Davies initial
conditions for the large $k$-mode functions. The relation between
$\Phi_k$ and $v$ and the  
initial conditions on $v$ lead at once to a determination of the
coefficients $a_k$ and $b_k$ for $k >> H_0$\cite{mukhanov} 
\begin{equation}
a_k = -\frac{3}{2} \left[\frac{8\pi}{3M^2_{Pl}}\right] \dot{\phi}_{eff}(0)
\sqrt{\frac{\pi}{2H_0}} \frac{1}{k} \; ; \; b_k = 0 .
\label{coeffs}
\end{equation} 

Thus we find that the amplitude of scalar metric perturbations after 
horizon crossing is given by
\begin{equation}
|\delta_k(t)| = k^{\frac{3}{2}}|\Phi_k(t)| \approx
\frac{3}{2} \left[\frac{8\sqrt{\pi}}{3M^2_{Pl}}\right] \dot{\phi}_{eff}(0)
\left(\frac{2H_0}{k}\right)^{\nu -\frac{3}{2}}
e^{(2\nu-3)H_0t}.
\label{perts}
\end{equation}
The power spectrum per logarithmic $k$  interval is given by $|\delta_k(t)|^2$. The time dependence of $|\delta_k|$ displays the unstable growth associated with the spinodal instabilities of super-horizon modes and
is a hallmark of the phase transition.
This time dependence can be also understood from the constraint equation
that relates the Bardeen potential to the gauge invariant field fluctuations\cite{mukhanov}, which in longitudinal gauge are identified
with $\delta \phi(\vec x,t)$. 
 To obtain the amplitude and spectrum
of density perturbations at {\em second} horizon crossing we use the
conservation law associated with the gauge invariant variable\cite{mukhanov}
\begin{equation}
\xi_k = \frac{\frac{2\dot{\Phi}_k}{3H}+\Phi_k}{1+p/\varepsilon} + \Phi_k
\; \; ; \; \; \dot{\xi}_k =0 ,
\end{equation}
which is valid after horizon crossing of the mode with wavevector k. Using
this conservation law, and the relation that during the inflationary stage
$1+p/\varepsilon= 8\pi \dot{\phi}^2_{eff}/2M^2_{Pl}H^2_0 \ll 1$, and assuming matter domination at second horizon crossing  and $\dot{\Phi}_k(t_f)=0$\cite{mukhanov}, we find
\begin{equation}
|\delta_k(t_f)| = \frac{12 \Gamma(\nu)\sqrt{\pi}}{5 (\nu-\frac{3}{2})
{\cal F}(H_0/m)} \left(\frac{2H_0}{k}\right)^{\nu-\frac{3}{2}} ,
\label{amplitude}
\end{equation}
where ${\cal F}(H_0/m)$ determines the initial amplitude of the effective
zero mode (\ref{effzeromodein}). 
We can now read the power spectrum per logarithmic $k$ interval
\begin{equation}
{\cal P}_s(k) = |\delta_k|^2 \propto k^{-2(\nu-\frac{3}{2})},
\end{equation}
leading to the index for scalar density perturbations
\begin{equation}
n_s = 1-2(\nu-\frac{3}{2}). \label{index}
\end{equation}
We remark that we have not included the small corrections to the dynamics
of the effective zero mode and the scale factor arising from the non-linearities. These are expected to lead to perturbatively small ${\cal O}(\lambda)$ corrections to the index (\ref{index}). 
The spectrum given by (\ref{amplitude}) is
similar to that obtained in references\cite{turner,guthpi} although
the amplitude differs from that obtained there. We emphasize an important
feature of the spectrum: it has more power at {\em long wavelengths} because $\nu-3/2 > 0$. This is recognized to be a consequence
of the spinodal instabilities that result in the growth of long wavelength
modes and therefore in more power for these modes. 
This seems to be a robust prediction of new inflationary scenarios in
which the potential has negative second derivative in the region of field
space that produces inflation.  

 It is at this
stage that we recognize the consistency of our approach for separating
the composite effective zero mode from the small fluctuations. We have
argued above that many more than 60 e-folds are required for consistency,
and that the `small fluctuations' correspond to those modes that cross
the horizon during the last 60 e-folds of the inflationary stage. For these
modes $H_0/k = e^{-H_0 t^*(k)}$ where $t^*(k)$ is the time of horizon
crossing of the mode with wavevector $k$ since the beginning of inflation. 
The scale that  corresponds to the Hubble radius today $\lambda_0 =2\pi/k_0$ is the first to cross during the last 60 or so e-folds before the end of
inflation. Smaller scales today will correspond to $k > k_0$ at the
onset of inflation since they will cross the first horizon later and therefore will reenter earlier. The bound on $|\delta_{k_0}| \propto 
\Delta T/ T \leq  10^{-5}$ on
these scales provides a lower bound on the number of e-folds required for
these type of models to be consistent:
\begin{equation}
N_e >
60+\frac{12}{\nu-\frac{3}{2}}-
\frac{\ln(\nu-\frac{3}{2})}{\nu-\frac{3}{2}}, \label{numbofefolds}
\end{equation}
where we have written the total number of e-folds as $N_e=H_0\, t^*(k_0)+60$.
This in turn can be translated into an upper bound on the coupling
constant using the estimate given by eq.(\ref{efolds}).

\section{Contact with the Reconstruction Program:}

The program of reconstruction of the inflationary potential seeks to
establish a relationship between features of the inflationary scalar
potential and the spectrum of scalar and tensor perturbations.
This program, in combination with measurements of scalar and tensor
components either from refined measurements of temperature inhomogeneities
of the CMB or through galaxy correlation functions will then offer a glimpse of the possible realization of the inflation\cite{reconstruction,lyth}. 
Such a reconstruction program is based on the slow roll approximation
and the spectral index of scalar and tensor perturbations are obtained
in a perturbative expansion in the slow roll
parameters\cite{reconstruction,lyth} 
\begin{eqnarray}
\epsilon(\phi) & = &
\frac{\frac{3}{2}\dot{\phi}^2}{\frac{\dot{\phi}^2}{2}+V(\phi)}\; ,
\label{epsifi}  \\
\eta(\phi) & = & -\frac{\ddot{\phi}}{H \dot{\phi}}\; . \label{etafi}
\end{eqnarray}
We can make contact with the reconstruction program by identifying $\phi$
above with our $\phi_{eff}$ after the first few e-folds of inflation needed
to assemble the effective zero mode from the quantum
fluctuations. We have numerically established that for the weak scalar
coupling required 
for the consistency of these models, the cosmologically interesting scales
cross the horizon during the epoch in which $H \approx H_0 \; ; \;
\dot{\phi}_{eff} \approx (\nu - 3/2)\; H_0 \; \phi_{eff} \; ; \; V \approx
m_R^4/\lambda \gg \dot{\phi}^2_{eff}$. In this case we find 
\begin{equation}
\eta(\phi_{eff}) = \nu - \frac{3}{2} \; ; \;  \epsilon(\phi_{eff}) \approx
{\cal O}(\lambda) \ll  \eta(\phi_{eff}).
\end{equation}

With these identifications, and in the notation of\cite{reconstruction,lyth}
the reconstruction program predicts  the index for scalar density
perturbations $n_s$ given by
\begin{equation}
 n_s-1 = -2(\nu - \frac{3}{2})+ {\cal O}(\lambda),
\end{equation}
which coincides with the index for the spectrum given by the power
spectrum per logarithmic interval $|\delta_k|^2$ with $|\delta_k|$
given by eq.(\ref{amplitude}).  
We must note however that our treatment did not
assume slow roll for which $(\nu - \frac{3}{2})\ll 1$. Our
self-consistent, non-perturbative study of the dynamics plus the
underlying 
requirements for the identification of a composite operator acting as an
effective zero mode, validates the reconstruction program in weakly
coupled new inflationary models.

\section{Conclusions:} 
We have studied the non-equilibrium dynamics of a new inflation scenario
in a self-consistent, non-perturbative framework based on a large $N$
expansion, 
including the dynamics of the scale factor and backreaction of quantum
fluctuations. Quantum fluctuations associated with superhorizon modes
grow exponentially as a result of the spinodal instabilities and
contribute to the 
energy momentum tensor in such a way as to end inflation consistently.

Analytical and numerical estimates have been provided that
establish the regime of validity of the classical approach.
 We find that these  superhorizon modes re-assemble into an
effective zero mode and unambiguously identify the 
composite operator that can be used as an effective expectation value of
the inflaton field whose {\em classical} dynamics drives the evolution of
the scale factor. This identification also provides the initial condition
for this effective zero mode. 

If the model allows many more than 60 e-folds we provide a criterion that
allows the identification of the small perturbations that give rise to scalar
metric (curvature) perturbations. We then use this criterion combined with
the gauge invariant approach  to obtain the spectrum for scalar perturbations.
We find the index to be less than one, providing more power at long wavelength
as a result of the spinodal instabilities. We argue that this `red' spectrum
is a robust feature of potentials that lead to spinodal instabilities in the
region in field space associated with inflation. Finally we made contact with
the reconstruction program and validated the results for these type of models
based on the slow-roll assumption, despite the fact that our study does
not involve such an approximation. A more detailed version of this
article with a discussion of the issue of decoherence is
forthcoming\cite{next}. 
\section{Acknowledgements:} 
The authors thank 
J. Baacke, A. Dolgov, L. Grishchuk, K. Heitman,  E. Kolb,  D. Polarski
and E. Weinberg for conversations and discussions. D. B. thanks the
N.S.F for partial support through the grant 
awards: PHY-9605186 and INT-9216755, the Pittsburgh Supercomputer Center for
grant award No: PHY950011P and LPTHE for warm hospitality.  R. H., 
D. C. and S. P. K. were supported by DOE grant DE-FG02-91-ER40682.

\begin{figure}
\epsfig{file=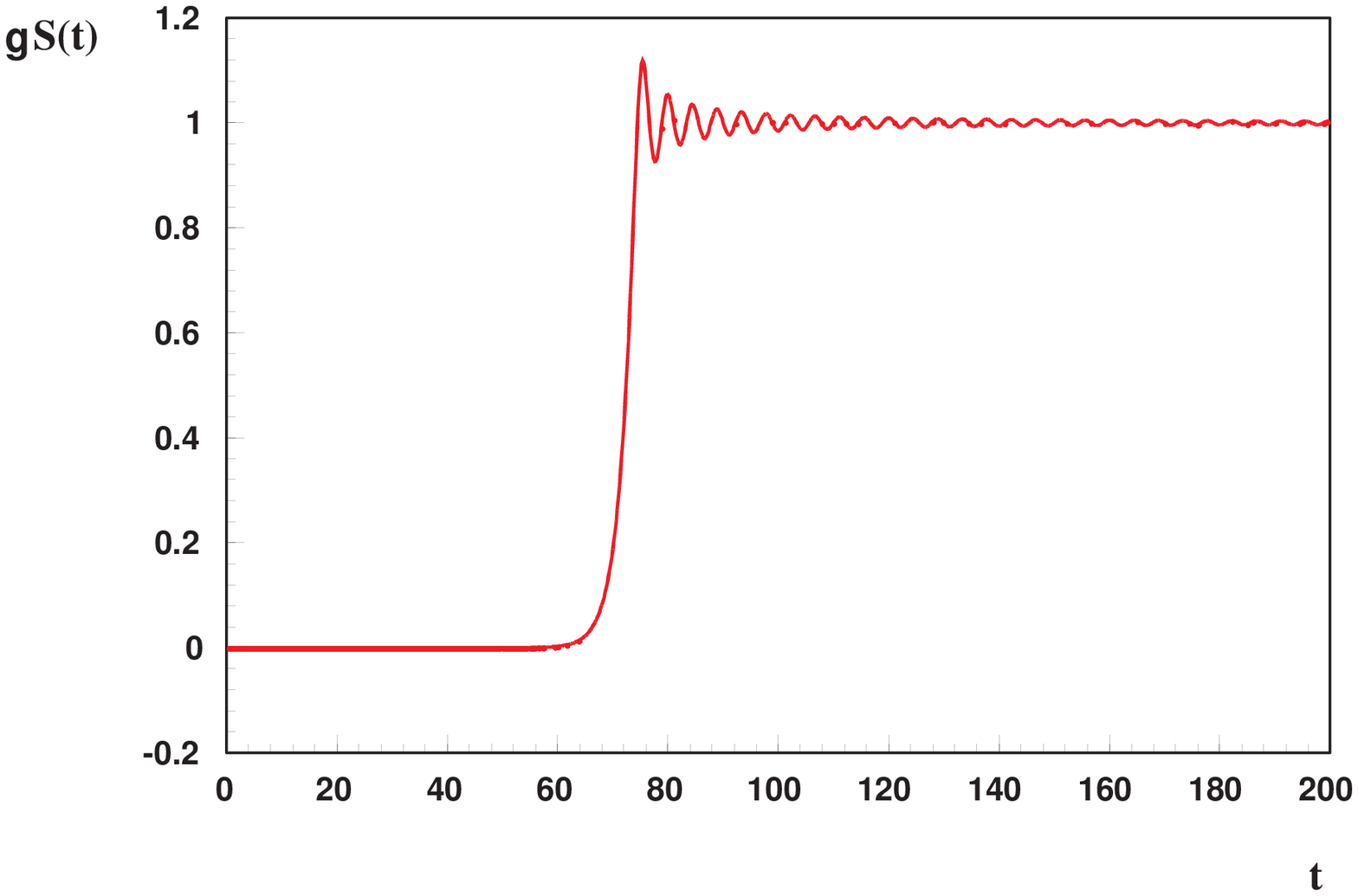,width=5.5in,height=3.2in}
\caption{ $g\Sigma$ vs. $\tau$, for $\eta(0)=0, \dot{\eta}(0)=0,
 \lambda= 10^{-12}, h_0 = 1.5$ }
\end{figure}

\begin{figure}
\epsfig{file=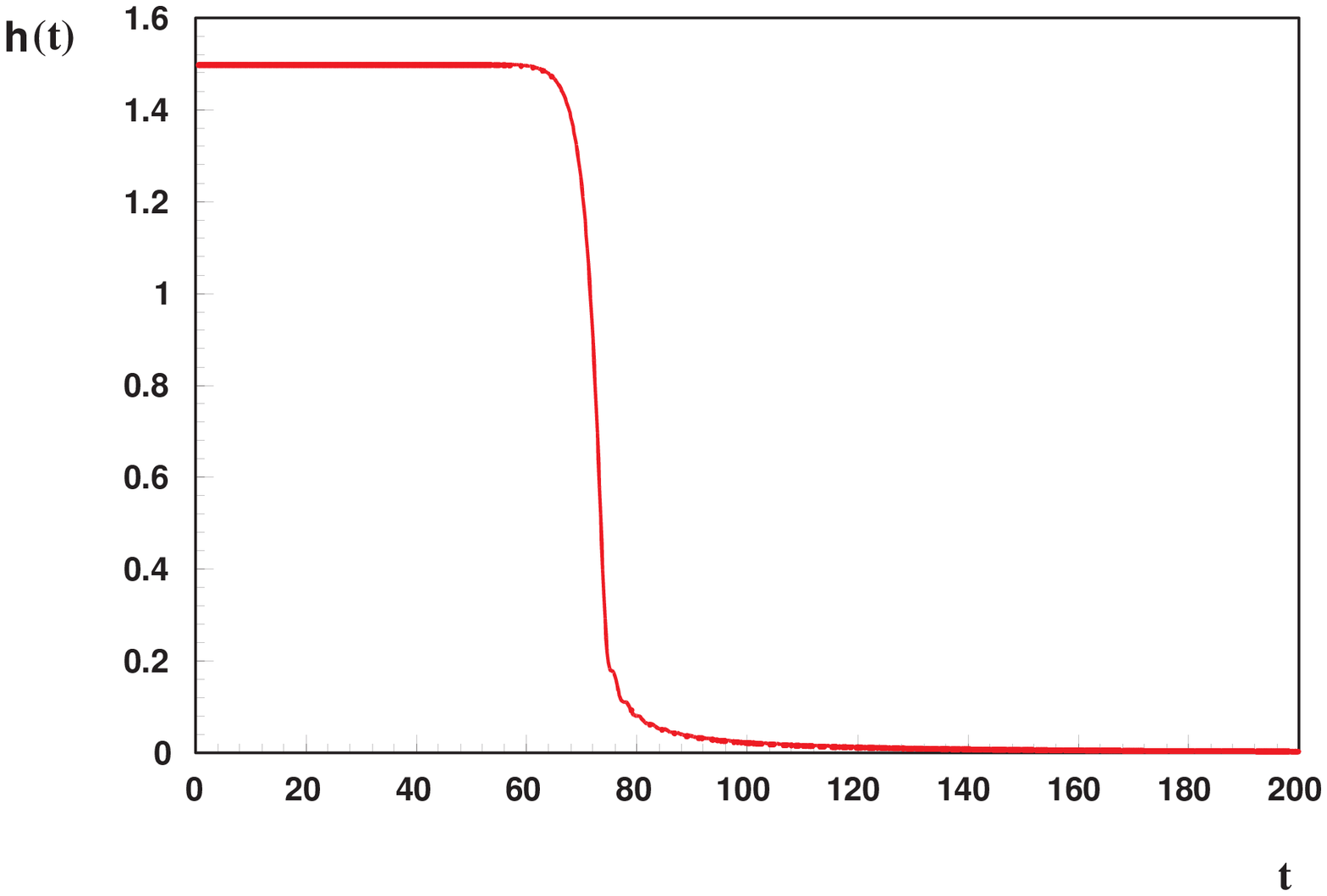,width=5.5in,height=3.2in}
\caption{$H(\tau)$ vs. $\tau$, for $\eta(0)=0, \dot{\eta}(0)=0, 
 \lambda = 10^{-12}, h_0 = 1.5 $ }
\end{figure}

\begin{figure}
\epsfig{file=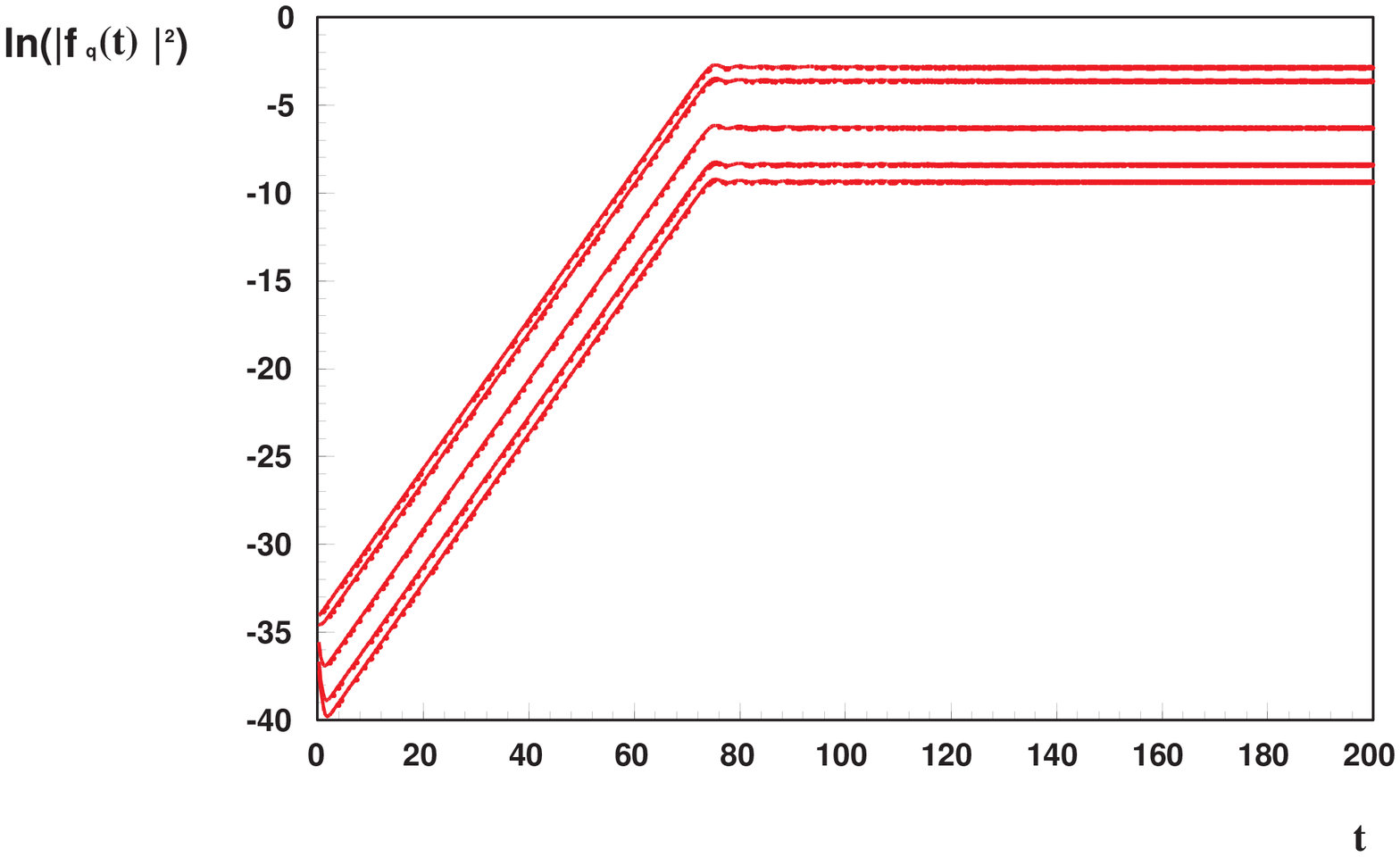,width=5.5in,height=3.2in}
\caption{$\ln(|f_q(\tau)|^2)$ vs. $\tau$, for $\eta(0)=0, \dot{\eta}(0)=0,  \lambda= 10^{-12}, h_0=1.5$ for $q=0.5,2.0,5.0,8.0,10$ smaller
$ q $ corresponds to larger values}
\end{figure}

\begin{figure}
\epsfig{file=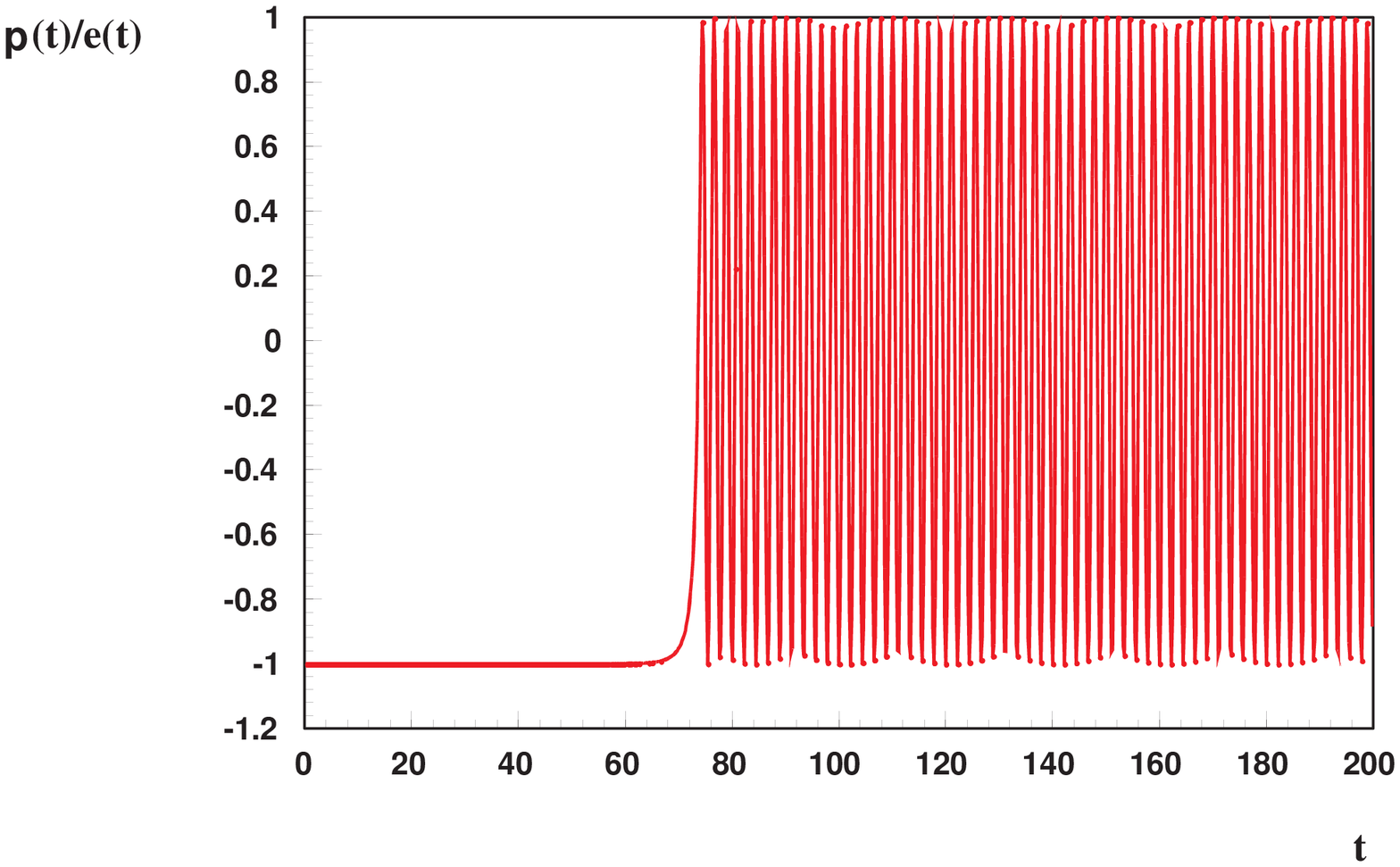,width=5.5in,height=3.2in}
\caption{$p/\varepsilon$ vs. $\tau$, for $\eta(0)=0, \dot{\eta}(0)=0,
 \lambda= 10^{-12}, h_0=1.5$ }
\end{figure}

\begin{figure}
\epsfig{file=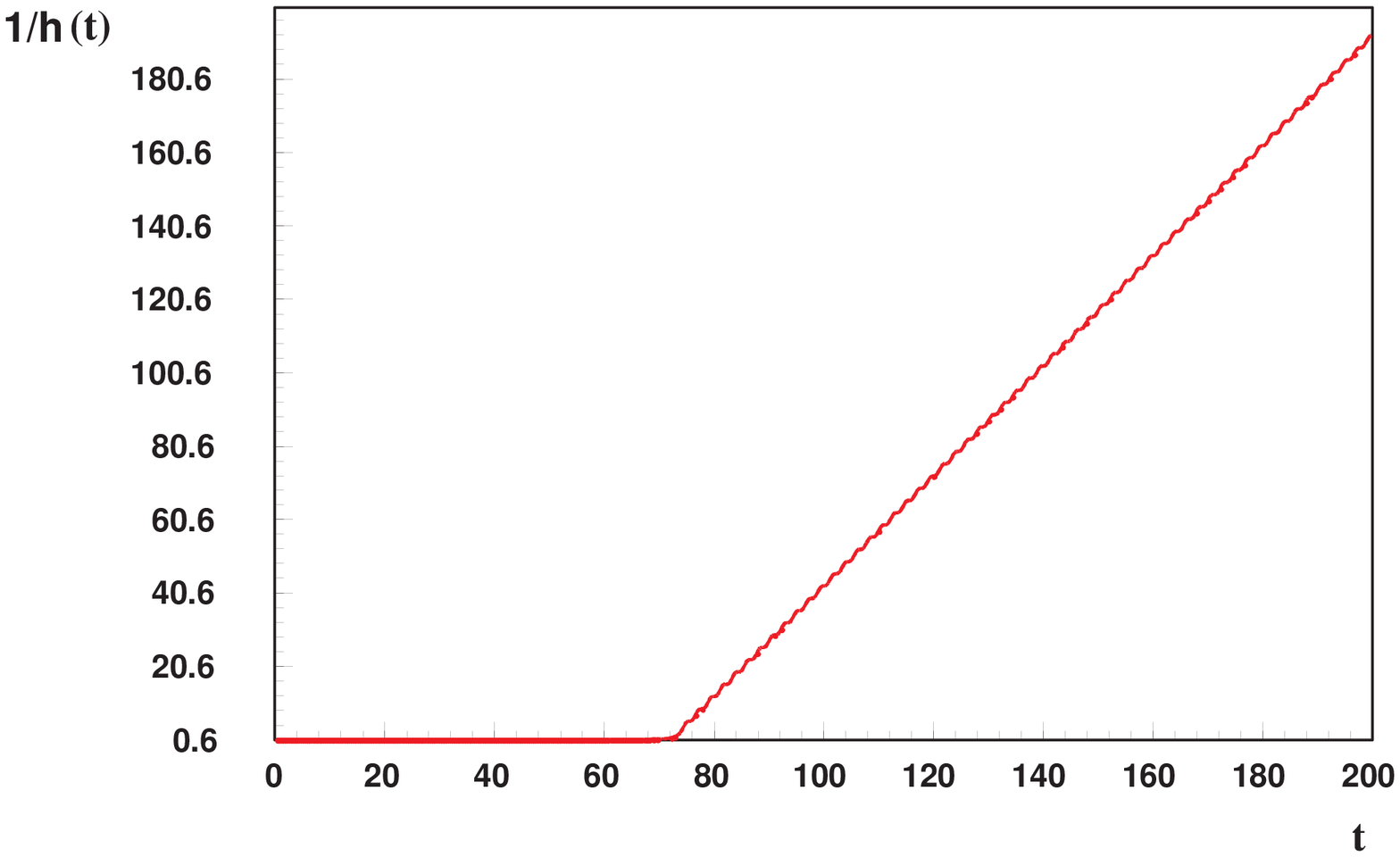,width=5.5in,height=3.2in}
\caption{$1/h(\tau)$ vs. $\tau$, for $\eta(0)=0, \dot{\eta}(0)=0,
 \lambda= 10^{-12}, h_0=1.5$ }
\end{figure}
\end{document}